\pdfoutput=1
\documentclass[cameraready]{Interspeech}

\usepackage{cite}
\usepackage{amsmath,amssymb,amsfonts}
\usepackage{algorithmic}
\usepackage{graphicx}
\usepackage{xcolor}
\usepackage{booktabs}
\usepackage{enumitem}
\newcommand{\para}[1]{\noindent\textbf{#1}}

\title{ParaSpeechCLAP: A Dual-Encoder Speech-Text Model for
\texorpdfstring{\\}{ } Rich Stylistic Language-Audio Pretraining}

\author[affiliation={1}, orcid=0000-0002-1783-3717]{Anuj}{Diwan}
\author[affiliation={2}, orcid=0000-0003-3607-9104]{Eunsol}{Choi}
\author[affiliation={1}, orcid=0009-0008-4465-1987]{David}{Harwath}
\address{
    $^1$ Department of Computer Science, The University of Texas at Austin, USA \\
    $^2$ Computer Science and Data Science, New York University, USA
}

\email{anuj.diwan@utexas.edu, eunsol@nyu.edu, harwath@utexas.edu}


\usepackage{comment}

\begin{document}

\maketitle
% the abstract here must exactly match the abstract entered into the paper submission system
\begin{abstract}
% 1000 characters. ASCII characters only. No citations.
We introduce ParaSpeechCLAP, a family of dual-encoder models that map speech and text style captions into a shared embedding space, supporting rich intrinsic (speaker-level) and situational (utterance-level) descriptors, such as pitch, texture, and emotion, beyond the narrow set handled by existing models. We train separate Intrinsic and Situational models alongside a unified Combined model, finding that specialized models are stronger on individual style dimensions while the unified model excels on compositional evaluation. We further show that ParaSpeechCLAP-Intrinsic benefits from an additional classification loss and class-balanced training. We demonstrate performance on style caption retrieval, speech attribute classification, and usability as inference-time reward models for style-prompted TTS. ParaSpeechCLAP models outperform baselines on most metrics across all three applications. Our models and code are released at \url{https://github.com/ajd12342/paraspeechclap}.
\end{abstract}

\section{Introduction}
Existing speech-caption alignment models~\cite{jing2024paraclapgenerallanguageaudio} handle only a narrow set of stylistic attributes. Yet, real-world speech varies along many more dimensions: pitch, texture, clarity, and beyond~\cite{diwan2025scalingrichstylepromptedtexttospeech}. While emotion recognition has made significant progress~\cite{ma2023emotion2vecselfsupervisedpretrainingspeech,huang2025dynamicsuperbphase2collaborativelyexpanding,xu2023secapspeechemotioncaptioning}, support for this broader set of speech styles, especially when specified as freeform natural-language descriptions, remains lacking. Closing this gap would benefit a range of applications, including style-prompted text-to-speech (TTS)~\cite{guo2022promptttscontrollabletexttospeechtext,lacombe-etal-2024-parler-tts,diwan2025scalingrichstylepromptedtexttospeech}, expressive speech retrieval~\cite{8003425,jing2024paraclapgenerallanguageaudio}, speech style captioning~\cite{yamauchi2023stylecapautomaticspeakingstylecaptioning,ando2024factorconditionedspeakingstylecaptioning}, and expressive spoken dialog systems~\cite{matsuura2025emonewsspokendialogueexpressive,papangelis2017ldsdsexpressivespokendialogue}.

This paper introduces ParaSpeechCLAP, a family of CLAP~\cite{elizalde2022claplearningaudioconcepts}-style dual-encoder models designed to map speech and rich textual style descriptions into a common embedding space. Following the ParaSpeechCaps~\cite{diwan2025scalingrichstylepromptedtexttospeech} taxonomy, we train three models: ParaSpeechCLAP-Intrinsic for speaker-level tags (e.g., pitch, texture, clarity) and ParaSpeechCLAP-Situational for utterance-level tags (e.g. emotion), and ParaSpeechCLAP-Combined trained on both tag types. We find that specialization yields stronger performance on individual style dimensions, while ParaSpeechCLAP-Combined excels on compositional evaluation that requires joint knowledge of both intrinsic and situational tags, suggesting the two strategies are complementary. Our models are among the first dual-encoder models to support intrinsic tags and a wider range of situational tags. We further demonstrate the use of such models for inference-time reward guidance to improve the style consistency of style-prompted TTS models in a training-free manner. While best-of-N selection with a learned scoring function has been explored in language modeling~\cite{stiennon2020learning} and image generation~\cite{radford2021learning}, to our knowledge this is the first application of a reward model for style-guided TTS selection.

Our approach builds upon the dual-encoder paradigm popularized by CLIP~\cite{radford2021learning} for image-text tasks and extended to general audio by CLAP~\cite{elizalde2022claplearningaudioconcepts}. In the speech domain, related methods have focused on other modalities or limited style sets; SpeechCLIP~\cite{shih2022speechclipintegratingspeechpretrained} aligns speech with images rather than text style descriptions, while ParaCLAP~\cite{jing2024paraclapgenerallanguageaudio} and SSE~\cite{10888883} handle only a small set of situational emotion tags. Unlike these, ParaSpeechCLAP models are trained on diverse, rich style captions that span both intrinsic and situational attributes. Furthermore, we find that ParaSpeechCLAP-Intrinsic benefits from a multitask contrastive plus classification loss that allows it to also predict specific style attributes. While ParaSpeechCLAP-Situational and ParaSpeechCLAP-Combined do not introduce architectural novelties beyond the encoder upgrades, we include them to provide complete coverage of the ParaSpeechCaps rich style tag taxonomy and to enable direct comparison between specialized and unified training strategies.

We validate our models on three downstream applications: style caption retrieval, rich speech attribute classification, and inference-time guidance for TTS. For the latter, we perform best-of-N selection by scoring multiple generated candidates against a target style caption, effectively guiding the TTS system toward more stylistically faithful output. In summary:
\begin{itemize}[noitemsep,nolistsep,leftmargin=*]
\item We introduce ParaSpeechCLAP, a family of dual-encoder models that learn a joint embedding space for speech and rich textual style descriptions spanning both intrinsic and situational attributes, making it one of the first model families to support this breadth of tags. We compare specialized and unified training strategies, finding them to be complementary.
\item We propose a classification loss for ParaSpeechCLAP-Intrinsic that uses the text encoder to produce class embeddings, and show it improves performance over a contrastive-only objective.
\item We demonstrate a novel application of dual-encoder models as inference-time reward models for style-prompted TTS through best-of-N selection, improving style consistency without any additional training.
\end{itemize}

\begin{figure*}
\centerline{\includegraphics[width=\linewidth]{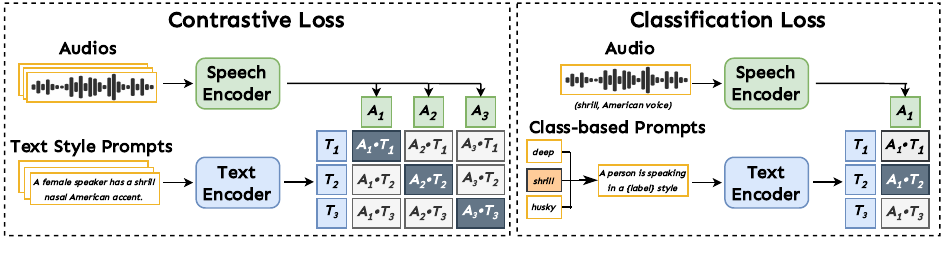}}
\caption{An overview of the ParaSpeechCLAP dual-encoder training methodology. (Left) All ParaSpeechCLAP models are trained with an InfoNCE contrastive loss to align speech and text embeddings. (Right) For ParaSpeechCLAP-Intrinsic, this is supplemented with a classification loss, where audio embeddings are aligned with text embeddings generated from templates filled with class labels.}
\label{fig:mainfig}
\vspace{-10pt}
\end{figure*}

\section{Methodology}
\label{sec:method}
As depicted in Figure~\ref{fig:mainfig}, each ParaSpeechCLAP model consists of speech and text encoders that project raw speech and text captions into a common multimodal embedding space. Their architecture is similar to that of the Contrastive Language-Audio Pretraining (CLAP)~\cite{elizalde2022claplearningaudioconcepts}-based ParaCLAP~\cite{jing2024paraclapgenerallanguageaudio} model, but we use more modern and powerful encoders (detailed in our experimental section) and train with a dataset of rich speech style captions, ParaSpeechCaps~\cite{diwan2025scalingrichstylepromptedtexttospeech}. We train ParaSpeechCLAP-Intrinsic and ParaSpeechCLAP-Situational as two specialized models and additionally train a unified model, ParaSpeechCLAP-Combined, on the combined dataset. For ParaSpeechCLAP-Intrinsic, we use a multitask objective combining contrastive and classification ($\mathcal{L}_{\text{contrastive}} + \mathcal{L}_{\text{classify}}$), while for ParaSpeechCLAP-Situational and ParaSpeechCLAP-Combined, we use only $\mathcal{L}_{\text{contrastive}}$; these losses are defined in Section~\ref{sec:contrastiveloss} and Section~\ref{sec:classificationloss}. Although the three models differ in their tag focus and training objectives, they share the same encoder architecture and embedding space, enabling direct comparison of specialization versus unification strategies.

\subsection{Training Example Format}
Each training example is a triplet $\{ (A_i, T_i, Y_i) \}_{i=1}^N$: a speech clip $A_i$, a text caption $T_i$, and a multi-hot tag vector $Y_i \in \{0,1\}^M$, where $Y_{ik}=1$ if $T_i$ expresses tag $k$. For example, for the text caption \textit{``A man speaks in a guttural tone with a British accent"}, the \textit{guttural} and \textit{British} entries in the tag vector are $1$ while all others are $0$. $Y_i$ is only used for ParaSpeechCLAP-Intrinsic's classification loss, for which $M=28$.

\subsection{Encoder Architecture}
Each model consists of a speech encoder $f_A(\cdot)$ and a text encoder $f_T(\cdot)$. Both encoders consist of a transformer backbone and a projection head (two linear transformations with a GELU activation and layer normalization) that map speech waveforms and text captions to a common $768$-dimensional embedding space. We use the $317$M-parameter WavLM-Large~\cite{chen2022wavlm} (chosen for its strong SUPERB benchmark~\cite{yang2021superbspeechprocessinguniversal} performance) as the speech encoder's backbone and compute a speech embedding by mean-pooling the last layer's hidden states and applying the projection head. We use Granite Embedding 278M Multilingual~\cite{awasthy2025graniteembeddingmodels} (chosen for its MTEB benchmark~\cite{muennighoff2023mtebmassivetextembedding} ranking among $300$M-parameter models) as the text encoder's backbone and compute a text embedding by taking the final-layer CLS token embedding and applying the projection head.

% To generate embeddings for the classification task, each tag in the vocabulary $\{C_k\}_{k=1}^M$ is first converted into a descriptive text prompt using a template (we use \texttt{A person is speaking in a \{tag\} style.}). These templatized tags are then passed through the same text encoder and projection head to produce a set of $M$ tag embeddings:
% \begin{equation}
%     E_k = \text{proj}_T(f_T(\text{template}(C_k)))
% \end{equation}

\begin{table*}[ht!]
\centering
% Updated column definition: Added space and 3 new columns (ccc) at the end
\setlength{\tabcolsep}{2pt} % Reduced slightly to accommodate the extra width
\begin{tabular}{l ccccc@{\hspace{1em}}ccccc@{\hspace{1em}}ccc}
\toprule
& \multicolumn{5}{c}{\textbf{Situational}} & \multicolumn{5}{c}{\textbf{Intrinsic}} & \multicolumn{3}{c}{\textbf{Combined}} \\
\cmidrule(lr){2-6} \cmidrule(lr){7-11} \cmidrule(lr){12-14}
\textbf{Model} & \textbf{R@1} & \textbf{R@10} & \textbf{MedR $\downarrow$} & \textbf{UAR} & \textbf{F1} & \textbf{R@1} & \textbf{R@10} & \textbf{MedR $\downarrow$} & \textbf{UAR} & \textbf{F1} & \textbf{R@1} & \textbf{R@10} & \textbf{MedR $\downarrow$} \\
\midrule
\multicolumn{14}{l}{\small{\textit{Dual-Encoder Baselines}}} \\
Random Projection & $0.14$ & $2.65$ & $170$ & $6.56$ & $2.66$ & $1.45$ & $17.66$ & $28$ & $23.64$ & $16.21$ & $0.00$ & $0.49$ & $725$ \\
ParaCLAP~\cite{jing2024paraclapgenerallanguageaudio} & $0.41$ & $3.98$ & $106$ & $11.56$ & $3.67$ & $1.95$ & $26.32$ & $19$ & $29.16$ & $20.65$ & $0.34$ & $1.67$ & $462$ \\
ParaCLAP-PSC & $15.64$ & $72.55$ & $5$ & $23.91$ & $23.57$ & $11.49$ & $60.69$ & $8$ & $40.27$ & $37.27$ & $5.58$ & $32.12$ & $23$ \\
% Expr Speech Retrieval~\cite{kang2025expressivespeechretrievalusing} &  &  & & & & & & & & & & & \\
\midrule
\multicolumn{14}{l}{\small{\textit{Classifier Baseline}}} \\
VoxProfile-VQ~\cite{feng2025voxprofilespeechfoundationmodel}  & - & - & - & - & - & - & - & - & $41.40$ & $\mathbf{40.24}$ & - & - & - \\
\midrule
\textbf{PSCLAP-Situational} & $24.79$ & $\mathbf{88.82}$ & $\mathbf{3}$ & $\mathbf{56.30}$ & $\mathbf{58.01}$ & $5.35$ & $37.67$ & $15$ & $31.08$ & $25.05$ & $12.71$ & $50.48$ & $\mathbf{10}$ \\
\textbf{PSCLAP-Intrinsic} & $4.54$ & $27.93$ & $26$ & $5.73$ & $3.08$ & $\mathbf{18.62}$ & $\mathbf{65.23}$ & $\mathbf{6}$ & $\mathbf{46.58}$ & $38.27$ & $1.53$ & $8.24$ & $163$ \\
\textbf{PSCLAP-Combined} & $\mathbf{25.62}$ & $83.58$ & $\mathbf{3}$ & $\mathbf{56.43}$ & $55.60$ & $13.51$ & $58.14$ & $7$ & $32.83$ & $31.74$ & $\mathbf{14.31}$ & $\mathbf{52.17}$ & $\mathbf{10}$ \\
\bottomrule
\end{tabular}%
\caption{Performance on situational, intrinsic, and combined evaluation datasets. We report retrieval metrics (Recall@k, Median Rank) and classification metrics (Unweighted Average Recall, Macro F1). For all metrics except median rank, higher is better. PSCLAP refers to ParaSpeechCLAP. Our ParaSpeechCLAP models outperform all baselines on most metrics. The unified ParaSpeechCLAP-Combined model is weaker than the specialized variants on specialized evaluation datasets and stronger on the combined evaluation dataset.}
\label{tab:combined_results}
\vspace{-15pt}
\end{table*}

\subsection{Contrastive Loss}
\label{sec:contrastiveloss}
The contrastive objective aligns speech clips with their corresponding text prompts using a standard bidirectional InfoNCE loss~\cite{oord2019representationlearningcontrastivepredictive}. We compute an $N \times N$ cosine similarity matrix $S$ between batch speech embeddings and text prompt embeddings, where $S_{ij} = \frac{f_A(A_i)^{\mathsf{T}} \cdot f_T(T_j)}{\lVert f_A(A_i)\rVert \lVert f_T(T_j)\rVert}$. The loss is computed bidirectionally with a learnable temperature $\tau$:
\begin{equation*}
\begin{split}
    \mathcal{L}_{\text{contrastive}} = \frac{1}{2N} \sum_{i=1}^N \Big( & -\log \frac{\exp(S_{ii} / \tau)}{\sum_{j} \exp(S_{ij} / \tau)} \\
    & -\log \frac{\exp(S_{ii} / \tau)}{\sum_{j} \exp(S_{ji} / \tau)} \Big)
\end{split}
\end{equation*}
\subsection{Inference-Like Classification Loss}
\label{sec:classificationloss}
Rather than training ParaSpeechCLAP-Intrinsic using a separate classification head, we use the text encoder itself to produce class embeddings that can then be used to create classification logits. The advantage of this approach over a dedicated head is that it keeps the text encoder active during classification training, reinforcing modality alignment. We first use Gemini 2.5 Pro~\cite{comanici2025gemini25pushingfrontier} to generate $6$ paraphrased captions for each of the $M=28$ rich intrinsic tags in the vocabulary. Then, for each minibatch, we freshly sample one caption for each tag and pass them through the text encoder to produce a set of $M$ tag embeddings $\{E_k\}|_{k=1}^{M}$, where $k$ is the class index. For each example, we compute classification logits by taking the speech embedding's dot product with these embeddings, $l_{ik} = f_A(A_i) \cdot E_k^\mathsf{T}$.
A binary cross-entropy loss is then applied using the ground-truth tag vector $Y_i$, where $\sigma(\cdot)$ is the sigmoid function:
\begin{equation*}
\begin{split}
    \mathcal{L}_{\text{classify}}^{(i)} = -\sum_{k=1}^M [ & Y_{ik} \log\sigma(l_{ik}) + (1 - Y_{ik}) \log(1 - \sigma(l_{ik}))]
\end{split}
\end{equation*}
The final classification loss is the average over the batch:
$ \mathcal{L}_{\text{classify}} = \frac{1}{N} \sum_{i=1}^N \mathcal{L}_{\text{classify}}^{(i)}$.

\section{Experimental Setup}
\subsection{Training}
\para{Dataset} We train our models on ParaSpeechCaps~\cite{diwan2025scalingrichstylepromptedtexttospeech}, a large-scale dataset that provides both manually and automatically annotated style prompts for speech clips from Expresso, EARS, and subsets of VoxCeleb and Emilia. We train ParaSpeechCLAP-Intrinsic and ParaSpeechCLAP-Situational on the intrinsic-tag and situational-tag subsets, respectively ($2412$ hr and $298$ hr), and train ParaSpeechCLAP-Combined on their union (upsampling the situational data to balance the two equally). We randomly truncate or pad each speech clip to a fixed $10$-second length.

\para{Hyperparameters} We train all models for $4500$ steps using the Adam optimizer with a constant learning rate of $10^{-5}$ on $4$ NVIDIA A40 GPUs with a per-GPU batch size of $32$. We train all model parameters including the backbone encoders. Our learnable temperature parameter is initialized to $0.07$. For ParaSpeechCLAP-Intrinsic, we additionally use class-balanced training by upsampling examples annotated with rare tags, using inverse tag sampling frequencies. ParaSpeechCLAP-Situational and Combined are trained without class-balanced sampling as we did not observe improvements.

\subsection{Evaluation}
We evaluate models for three applications: \textbf{retrieval}, \textbf{classification}, and \textbf{inference-time guidance for TTS} on situational, intrinsic, and combined-tag evaluation sets.

\para{Datasets} For retrieval and classification, we use the ParaSpeechCaps~\cite{diwan2025scalingrichstylepromptedtexttospeech} holdout set subdivided into three new evaluation sets: Situational, Intrinsic, and Combined. The Intrinsic set is drawn from the VoxCeleb portion of the ParaSpeechCaps holdout set and contains $2819$ speech clips and $64$ text style prompts with intrinsic-tag captions. The Situational and Combined sets are drawn from the Expresso-EARS portion of the ParaSpeechCaps holdout set, containing $1432$ speech clips each; the Situational set uses only situational-tag captions ($346$ in total), while the Combined set uses captions containing both situational and intrinsic tags ($1432$ in total). For TTS inference-time guidance, we use the tag-balanced test set from ParaSpeechCaps~\cite{diwan2025scalingrichstylepromptedtexttospeech}, consisting of $246$ examples with $5$ examples per tag. We evaluate on ParaSpeechCaps holdout and test sets because, to our knowledge, no existing benchmark supports the full breadth of rich style tags ($23$ situational and $28$ intrinsic) that the ParaSpeechCLAP model family supports. Standard situational benchmarks such as IEMOCAP~\cite{Busso2008} and RAVDESS~\cite{livingstone2018ryerson} cover only $5$--$6$ emotions.

\para{Retrieval Setup}
Given $N_s$ speech clips and $N_t$ text style prompts (where multiple clips may map to the same prompt), we compute embeddings for all of them, construct an $N_s \times N_t$ cosine similarity matrix, and rank text prompts per speech clip in descending order. We report:
\begin{itemize}[nolistsep,noitemsep,leftmargin=*]
    \item \textit{Recall@k (R@k)}: The percentage of speech clips for which the corresponding style prompt is ranked within the top $k \in \{1,10\}$ results.
    \item \textit{Median Rank (MedRank)}: The median rank of the ground-truth style prompt computed across all speech clips.
\end{itemize}

\para{Classification Setup}
For classification, we evaluate on speech clips that are each assigned to one of $C$ classes. Each class label is converted to a text prompt via the template \textit{A person is speaking in a \{label\} style}.\footnote{While classification performance may be sensitive to prompt wording, we leave its analysis to future work.} We compute cosine similarities between each speech clip's embedding and all $C$ prompts, apply softmax, and predict the highest-scoring class. We report:
\begin{itemize}[nolistsep,noitemsep,leftmargin=*]
    \item \textit{Unweighted Average Recall (UAR):} Recall is computed per class and then averaged across all classes.
    \item \textit{Macro F1-score (F1):} The F1-score (the harmonic mean of precision and recall) is calculated per class and then averaged across all classes.
\end{itemize}

\para{Inference-Time Guidance for TTS Setup}
We perform experiments on the best style-prompted TTS model from ParaSpeechCaps~\cite{diwan2025scalingrichstylepromptedtexttospeech}, which takes a style prompt and a transcript as input and generates speech. We generate $N=10$ candidate speech clips from the TTS model and compute cosine similarities between their ParaSpeechCLAP speech embeddings and the text embedding of the input style prompt, selecting the speech clip with the highest cosine similarity. We use ParaSpeechCLAP-Intrinsic for intrinsic-tag prompts and ParaSpeechCLAP-Situational for situational-tag prompts. We evaluate the selected clips using the same evaluation metrics and protocol as ParaSpeechCaps~\cite{diwan2025scalingrichstylepromptedtexttospeech}: Consistency MOS (CMOS), Intrinsic and Situational Tag Recalls, Naturalness MOS (NMOS), and WER. CMOS, NMOS, and tag recalls are obtained via human listening tests with three raters per sample and both AB and BA presentation orders to minimize first-option bias. CMOS and NMOS use a 5-point Likert scale. WER is computed automatically.

\section{Results and Discussion}
% This section presents the main quantitative results and initial discussion.

\subsection{Baselines}
For TTS guidance, we compare inference with and without ParaSpeechCLAP guidance. For retrieval and classification:
\begin{itemize}[nolistsep,noitemsep,leftmargin=*]
    \item \textit{Random Projection}: Our ParaSpeechCLAP architecture with pretrained encoder weights and random projector weights.
    \item \textit{ParaCLAP}~\cite{jing2024paraclapgenerallanguageaudio}: An existing speech-prompt model trained on MSP-Podcast, which contains a $6$-tag subset of ParaSpeechCaps' rich situational tags and none of its intrinsic tags.
    \item \textit{ParaCLAP-PSC}: The ParaCLAP model fine-tuned on ParaSpeechCaps. This baseline isolates the contribution of the ParaSpeechCaps data to our ParaSpeechCLAP models.
    % \item \textit{Expr Speech Retrieval}~\cite{kang2025expressivespeechretrievalusing}: A contemporaneous speech-prompt model trained on IEMOCAP, ESD and Expresso containing most of the ParaSpeechCaps rich situational tags and none of its rich intrinsic tags.
    \item \textit{VoxProfile-Voice Quality}~\cite{feng2025voxprofilespeechfoundationmodel}: An existing speech classifier (classification head over a WavLM~\cite{chen2022wavlm} encoder) trained on the human-annotated intrinsic-tag subset of ParaSpeechCaps. We only evaluate this model for intrinsic-tag classification as it cannot be used for situational tags or retrieval.
\end{itemize}

\subsection{Main Results}
\para{Retrieval and Classification}
We present results for the retrieval and classification tasks on situational, intrinsic, and combined evaluation datasets in Table~\ref{tab:combined_results}. Our ParaSpeechCLAP models significantly outperform existing dual-encoder baselines on all metrics. ParaCLAP-PSC improves over the original ParaCLAP, confirming that ParaSpeechCaps is a valuable training dataset, but ParaSpeechCLAP models outperform ParaCLAP-PSC, demonstrating that our encoder choice and training methodology improvements provide gains beyond what the data offers. Our models are competitive with the VoxProfile~\cite{feng2025voxprofilespeechfoundationmodel} classifier baseline, outperforming it on UAR ($46.58$ vs.\ $41.40$) while underperforming on macro F1 ($38.27$ vs.\ $40.24$), i.e., achieving better per-class recall but worse precision, suggesting that ParaSpeechCLAP-Intrinsic overpredicts certain frequent tags. We note that while the classifier baseline can only perform classification on a fixed set of intrinsic tags, our models support both intrinsic and situational tags and can perform both retrieval and classification using natural-language prompts.

The ParaSpeechCLAP-Combined model achieves the strongest results on the combined evaluation dataset (e.g., $14.31$ R@1 and $52.17$ R@10), outperforming specialized models.\footnote{We report only retrieval metrics for the combined evaluation set, as the compositional captions do not map to single class labels, making the classification setup inapplicable.} However, on the specialized evaluation sets, ParaSpeechCLAP-Combined underperforms the corresponding specialized model: it trails ParaSpeechCLAP-Intrinsic on intrinsic evaluation (e.g., $13.51$ vs.\ $18.62$ R@1) and is slightly weaker than or matches ParaSpeechCLAP-Situational on situational evaluation (e.g., $83.58$ vs.\ $88.82$ R@10). The mismatched specialized models also show cross-domain transfer, both substantially exceeding the Random Projection baseline on the other's evaluation set. Overall, a single model struggles to excel on all tag types, but ParaSpeechCLAP-Combined offers the best trade-off when both tag types are present simultaneously.

\para{Inference-Time Guidance for TTS}
We present results for ParaSpeechCLAP-guided best-of-N TTS inference in Table~\ref{tab:results}. We find that using ParaSpeechCLAP models for TTS guidance results in improved style consistency for both intrinsic and situational tags, demonstrating a novel use case of dual-encoder models as a training-free method to improve speech synthesis models. We also confirm that this selection method does not degrade naturalness (NMOS) and intelligibility (WER).

\begin{table}\centering
\setlength{\tabcolsep}{2pt}
\begin{tabular}{lccccc}\toprule
& \multicolumn{3}{c}{\textbf{Style Consistency}} & \multicolumn{1}{c}{\textbf{Quality}} & \multicolumn{1}{c}{\textbf{Intell.}} \\\cmidrule(lr){2-4}\cmidrule(lr){5-5}\cmidrule(lr){6-6}
\textbf{Model} & \textbf{CMOS} & \textbf{Int TR} & \textbf{Sit TR} & \textbf{NMOS} & \textbf{WER} $\mathbf{\downarrow}$ \\\midrule
GT & $4.04 \scriptstyle{\pm 0.07}$ & $73.1\%$ & $82.3\%$ & $4.42 \scriptstyle{\pm 0.06}$ & $8.04$ \\
\cmidrule{1-6}
Vanilla & $3.61 \scriptstyle{\pm 0.07}$ & $57.9\%$ & $69.2\%$ & $\mathbf{3.34} \scriptstyle{\pm 0.09}$ & $8.14$ \\
\textbf{w/ PSCLAP} & $\mathbf{3.70} \scriptstyle{\pm 0.07}$ & $\mathbf{62.4\%}$ & $\mathbf{74.3\%}$ & $\mathbf{3.35} \scriptstyle{\pm 0.09}$ & $\mathbf{7.56}$ \\
\bottomrule
\end{tabular}
\caption{Evaluation results comparing TTS inference without (Vanilla) and with ParaSpeechCLAP-guided best-of-N selection. We report style consistency (CMOS, Intrinsic Rich Tag Recall, and Situational Rich Tag Recall), speech quality (NMOS), and intelligibility (WER). PSCLAP refers to ParaSpeechCLAP. Using ParaSpeechCLAP models results in improved style consistency without affecting quality or intelligibility.}\label{tab:results}
\end{table}

\subsection{Ablation Study: ParaSpeechCLAP-Intrinsic}
\begin{table}
\centering
\setlength{\tabcolsep}{2pt}
\begin{tabular}{l ccccc}
\toprule
& \multicolumn{5}{c}{\textbf{Intrinsic}} \\
\cmidrule(lr){2-6}
\textbf{Model} & \textbf{R@1} & \textbf{R@10} & \textbf{MedR $\downarrow$} & \textbf{UAR} & \textbf{F1} \\
\midrule
\textbf{PSCLAP-Intrinsic} & $\mathbf{18.62}$ & $\mathbf{65.23}$ & $\mathbf{6}$ & $\mathbf{46.58}$ & $\mathbf{38.27}$ \\
\midrule
\multicolumn{6}{l}{\small{\textit{Ablations}}} \\
w/o New Encoders & $11.77$ & $63.63$ & $7$ & $41.44$ & $33.18$ \\
w/o Multitask & $13.76$ & $63.17$ & $7$ & $33.11$ & $30.00$ \\
w/o Class-Balancing & $13.94$ & $63.92$ & $\mathbf{6}$ & $40.60$ & $31.87$ \\
\bottomrule
\end{tabular}%
\caption{Ablation study for PSCLAP-Intrinsic. We compare our final model against versions with components removed.}
\vspace{-15pt}
\label{tab:ablation_results}
\end{table}
While ParaSpeechCLAP-Situational and ParaSpeechCLAP-Combined train well with a contrastive loss, we make additional design choices, such as a multitask classification loss and class balancing, to improve the performance of ParaSpeechCLAP-Intrinsic. To understand the contribution of each of these components, we conduct ablations that remove each one at a time, with all other training settings kept the same: \textit{w/o New Encoders} uses the original ParaCLAP~\cite{jing2024paraclapgenerallanguageaudio} encoders (a fine-tuned wav2vec2 and BERT), \textit{w/o Multitask} uses only the contrastive loss, and \textit{w/o Class-Balancing} removes class-balanced batch sampling. Table~\ref{tab:ablation_results} shows the results of these ablations. We find that removing any one of these components degrades performance, showing that each is important. Combined with the ParaCLAP-PSC baseline in Table~\ref{tab:combined_results}, which controls for training data, these ablations confirm that our gains stem from both the stronger encoders and the additional training techniques (multitask loss and class-balanced sampling), not from data scale alone. We focus ablations on ParaSpeechCLAP-Intrinsic, as it uses the most design components (multitask loss and class-balanced sampling); ParaSpeechCLAP-Situational uses only the contrastive loss with no additional techniques, leaving fewer components to ablate.

% \subsection{Discussion}
% \para{Prompt Augmentation Eval Set}\anuj{Optional: Add discussion}

% \para{Confusion Matrix}\anuj{Optional: Add discussion}
%   - Present and discuss a confusion matrix for zero-shot classification on a representative dataset (e.g., Expresso+EARS for situational, PSC-Intrinsic for intrinsic).
%   - Identify common confusions between tags.
%   - Discuss what this reveals about the model's understanding of different styles and potential ambiguities.
%   - Provide examples of correctly and incorrectly classified instances if space permits.

% \section{Related Work}
% Discuss existing work in several categories:
%   - CLAP (for audio-text): Briefly explain CLAP and its successes.
%   - CLIP (for image-text): Mention as an inspiration and successful similar paradigm.
%   - SpeechCLIP (for image-speech): Discuss its relevance and differences.
%   - Existing speech-text models for style/tags:
%       - ParaCLAP and SSE: Detail what they do, their focus (smaller set of situational tags), and how ParaSpeechCLAP differs/improves upon them.
% Emphasize the gap ParaSpeechCLAP is filling (intrinsic tags, broader situational tags, novel negative mining).

\section{Conclusion}
We introduced ParaSpeechCLAP, a family of dual-encoder models that create a shared embedding space for speech and rich textual style descriptions. Through extensive experiments, we demonstrated the ParaSpeechCLAP model family's ability to handle a diverse range of intrinsic and situational attributes for retrieval and classification, pioneered the use of its models as inference-time reward models for TTS systems, and ablated their design choices. A practical limitation of the specialized models is that they require selecting the appropriate variant at inference time; closing the gap between ParaSpeechCLAP-Combined and the specialized models remains an open challenge. Extending the best-of-N guidance strategy (which currently scales linearly with N in inference cost) to more efficient approaches such as guided decoding is also a promising future direction. We hope that ParaSpeechCLAP will encourage further research on rich style modeling and the development of rich style evaluation benchmarks.

% \bigskip
% \centerline{\large\bf Generative AI Usage Disclosure}
% \smallskip
\section{Generative AI Usage Disclosure}
The authors take full responsibility and are accountable for the contents of this paper. Generative AI tools were only used for light editing, polishing, and finding typos.

\bibliographystyle{IEEEtran}
\bibliography{references}

@article{livingstone2018ryerson,
  title={The Ryerson Audio-Visual Database of Emotional Speech and Song (RAVDESS): A dynamic, multimodal set of facial and vocal expressions in North American English},
  author={Livingstone, Steven R and Russo, Frank A},
  journal={PloS one},
  volume={13},
  number={5},
  pages={e0196391},
  year={2018},
  publisher={Public Library of Science}
}

@article{Busso2008,
  author  = {Busso, Carlos and Bulut, Murtaza and Lee, Chi-Chun and Kazemzadeh, Abe and Mower, Emily and Kim, Samuel and Chang, Jeannette N. and Lee, Sungbok and Narayanan, Shrikanth S.},
  title   = {{IEMOCAP}: interactive emotional dyadic motion capture database},
  journal = {Language Resources and Evaluation},
  year    = {2008},
  volume  = {42},
  number  = {4},
  pages   = {335--359},
  month   = dec,
}

@misc{oord2019representationlearningcontrastivepredictive,
      title={Representation Learning with Contrastive Predictive Coding}, 
      author={Aaron van den Oord and Yazhe Li and Oriol Vinyals},
      year={2019},
      eprint={1807.03748},
      archivePrefix={arXiv},
      primaryClass={cs.LG},
      url={https://arxiv.org/abs/1807.03748}, 
}

@inproceedings{muennighoff2023mtebmassivetextembedding,
    title = "{MTEB}: Massive Text Embedding Benchmark",
    author = "Muennighoff, Niklas  and
      Tazi, Nouamane  and
      Magne, Loic  and
      Reimers, Nils",
    editor = "Vlachos, Andreas  and
      Augenstein, Isabelle",
    booktitle = "Proceedings of the 17th Conference of the European Chapter of the Association for Computational Linguistics",
    month = may,
    year = "2023",
    address = "Dubrovnik, Croatia",
    publisher = "Association for Computational Linguistics",
    url = "https://aclanthology.org/2023.eacl-main.148/",
    doi = "10.18653/v1/2023.eacl-main.148",
    pages = "2014--2037"
}

@inproceedings{yang2021superbspeechprocessinguniversal,
  title     = {{SUPERB: Speech Processing Universal PERformance Benchmark}},
  author    = {Shu-wen Yang and Po-Han Chi and Yung-Sung Chuang and Cheng-I Jeff Lai and Kushal Lakhotia and Yist Y. Lin and Andy T. Liu and Jiatong Shi and Xuankai Chang and Guan-Ting Lin and Tzu-Hsien Huang and Wei-Cheng Tseng and Ko-tik Lee and Da-Rong Liu and Zili Huang and Shuyan Dong and Shang-Wen Li and Shinji Watanabe and Abdelrahman Mohamed and Hung-yi Lee},
  year      = {2021},
  booktitle = {{Interspeech 2021}},
  pages     = {1194--1198},
  doi       = {10.21437/Interspeech.2021-1775},
  issn      = {2958-1796},
}

@inproceedings{stiennon2020learning,
 author = {Stiennon, Nisan and Ouyang, Long and Wu, Jeffrey and Ziegler, Daniel and Lowe, Ryan and Voss, Chelsea and Radford, Alec and Amodei, Dario and Christiano, Paul F},
 booktitle = {Advances in Neural Information Processing Systems},
 editor = {H. Larochelle and M. Ranzato and R. Hadsell and M.F. Balcan and H. Lin},
 pages = {3008--3021},
 publisher = {Curran Associates, Inc.},
 title = {Learning to summarize with human feedback},
 url = {https://proceedings.neurips.cc/paper_files/paper/2020/file/1f89885d556929e98d3ef9b86448f951-Paper.pdf},
 volume = {33},
 year = {2020}
}

@INPROCEEDINGS{10888883,
  author={Zhang, Zixing and Wu, Yimeng and Dong, Zhongren and Xiang, Wulong and Shen, Shengfan and Schuller, Björn W.},
  booktitle={ICASSP 2025 - 2025 IEEE International Conference on Acoustics, Speech and Signal Processing (ICASSP)}, 
  title={SSE: A Speaking Style Extractor Based on Fine-Grained Contrastive Learning between Speech and Descriptive Text}, 
  year={2025},
  volume={},
  number={},
  pages={1-5},
  doi={10.1109/ICASSP49660.2025.10888883}}

@INPROCEEDINGS{shih2022speechclipintegratingspeechpretrained,
  author={Shih, Yi-Jen and Wang, Hsuan-Fu and Chang, Heng-Jui and Berry, Layne and Lee, Hung-yi and Harwath, David},
  booktitle={2022 IEEE Spoken Language Technology Workshop (SLT)}, 
  title={SpeechCLIP: Integrating Speech with Pre-Trained Vision and Language Model}, 
  year={2023},
  volume={},
  number={},
  pages={715-722},
  doi={10.1109/SLT54892.2023.10022954}}

@misc{comanici2025gemini25pushingfrontier,
      title={Gemini 2.5: Pushing the Frontier with Advanced Reasoning, Multimodality, Long Context, and Next Generation Agentic Capabilities}, 
      author={Gemini Team et. al.},
      year={2025},
      eprint={2507.06261},
      archivePrefix={arXiv},
      primaryClass={cs.CL},
      url={https://arxiv.org/abs/2507.06261}, 
}

@inproceedings{xu2023secapspeechemotioncaptioning,
  title={Secap: Speech emotion captioning with large language model},
  author={Xu, Yaoxun and Chen, Hangting and Yu, Jianwei and Huang, Qiaochu and Wu, Zhiyong and Zhang, Shi-Xiong and Li, Guangzhi and Luo, Yi and Gu, Rongzhi},
  booktitle={Proceedings of the AAAI Conference on Artificial Intelligence},
  volume={38},
  number={17},
  pages={19323--19331},
  year={2024}
}

@inproceedings{huang2025dynamicsuperbphase2collaborativelyexpanding,
title={Dynamic-{SUPERB} Phase-2: A Collaboratively Expanding Benchmark for Measuring the Capabilities of Spoken Language Models with 180 Tasks},
author={Chien-yu Huang et. al.},
booktitle={The Thirteenth International Conference on Learning Representations},
year={2025},
url={https://openreview.net/forum?id=s7lzZpAW7T}
}

@misc{papangelis2017ldsdsexpressivespokendialogue,
      title={LD-SDS: Towards an Expressive Spoken Dialogue System based on Linked-Data}, 
      author={Alexandros Papangelis and Panagiotis Papadakos and Margarita Kotti and Yannis Stylianou and Yannis Tzitzikas and Dimitris Plexousakis},
      year={2017},
      eprint={1710.02973},
      archivePrefix={arXiv},
      primaryClass={cs.IR},
      url={https://arxiv.org/abs/1710.02973}, 
}

@inproceedings{matsuura2025emonewsspokendialogueexpressive,
    title = "{E}mo{N}ews: A Spoken Dialogue System for Expressive News Conversations",
    author = "Matsuura, Ryuki  and
      Bharadwaj, Shikhar  and
      Liu, Jiarui  and
      Kunde Govindarajan, Dhatchinamoorthi",
    editor = "B{\'e}chet, Fr{\'e}d{\'e}ric  and
      Lef{\`e}vre, Fabrice  and
      Asher, Nicholas  and
      Kim, Seokhwan  and
      Merlin, Teva",
    booktitle = "Proceedings of the 26th Annual Meeting of the Special Interest Group on Discourse and Dialogue",
    month = aug,
    year = "2025",
    address = "Avignon, France",
    publisher = "Association for Computational Linguistics",
    url = "https://aclanthology.org/2025.sigdial-1.27/",
    pages = "338--342"
}

@misc{feng2025voxprofilespeechfoundationmodel,
      title={Vox-Profile: A Speech Foundation Model Benchmark for Characterizing Diverse Speaker and Speech Traits}, 
      author={Tiantian Feng and Jihwan Lee and Anfeng Xu and Yoonjeong Lee and Thanathai Lertpetchpun and Xuan Shi and Helin Wang and Thomas Thebaud and Laureano Moro-Velazquez and Dani Byrd and Najim Dehak and Shrikanth Narayanan},
      year={2025},
      eprint={2505.14648},
      archivePrefix={arXiv},
      primaryClass={cs.SD},
      url={https://arxiv.org/abs/2505.14648}, 
}

@INPROCEEDINGS{elizalde2022claplearningaudioconcepts,
  author={Elizalde, Benjamin and Deshmukh, Soham and Ismail, Mahmoud Al and Wang, Huaming},
  booktitle={ICASSP 2023 - 2023 IEEE International Conference on Acoustics, Speech and Signal Processing (ICASSP)}, 
  title={CLAP Learning Audio Concepts from Natural Language Supervision}, 
  year={2023},
  volume={},
  number={},
  pages={1-5},
  doi={10.1109/ICASSP49357.2023.10095889}}

@inproceedings{jing2024paraclapgenerallanguageaudio,
  title     = {ParaCLAP – Towards a general language-audio model for computational paralinguistic tasks},
  author    = {Xin Jing and Andreas Triantafyllopoulos and Björn Schuller},
  year      = {2024},
  booktitle = {Interspeech 2024},
  pages     = {1155--1159},
  doi       = {10.21437/Interspeech.2024-1315},
  issn      = {2958-1796},
}

@misc{awasthy2025graniteembeddingmodels,
      title={Granite Embedding Models}, 
      author={Parul Awasthy and Aashka Trivedi and Yulong Li and Mihaela Bornea and David Cox and Abraham Daniels and Martin Franz and Gabe Goodhart and Bhavani Iyer and Vishwajeet Kumar and Luis Lastras and Scott McCarley and Rudra Murthy and Vignesh P and Sara Rosenthal and Salim Roukos and Jaydeep Sen and Sukriti Sharma and Avirup Sil and Kate Soule and Arafat Sultan and Radu Florian},
      year={2025},
      eprint={2502.20204},
      archivePrefix={arXiv},
      primaryClass={cs.IR},
      url={https://arxiv.org/abs/2502.20204}, 
}

@inproceedings{diwan2025scalingrichstylepromptedtexttospeech,
    title = "Scaling Rich Style-Prompted Text-to-Speech Datasets",
    author = "Diwan, Anuj  and
      Zheng, Zhisheng  and
      Harwath, David  and
      Choi, Eunsol",
    editor = "Christodoulopoulos, Christos  and
      Chakraborty, Tanmoy  and
      Rose, Carolyn  and
      Peng, Violet",
    booktitle = "Proceedings of the 2025 Conference on Empirical Methods in Natural Language Processing",
    month = nov,
    year = "2025",
    address = "Suzhou, China",
    publisher = "Association for Computational Linguistics",
    url = "https://aclanthology.org/2025.emnlp-main.180/",
    doi = "10.18653/v1/2025.emnlp-main.180",
    pages = "3639--3659",
    ISBN = "979-8-89176-332-6"
}

@inproceedings{ando2024factorconditionedspeakingstylecaptioning,
  title     = {{Factor-Conditioned Speaking-Style Captioning}},
  author    = {Atsushi Ando and Takafumi Moriya and Shota Horiguchi and Ryo Masumura},
  year      = {2024},
  booktitle = {{Interspeech 2024}},
  pages     = {782--786},
  doi       = {10.21437/Interspeech.2024-633},
  issn      = {2958-1796},
}

@ARTICLE{8003425,
  author={Lotfian, Reza and Busso, Carlos},
  journal={IEEE Transactions on Affective Computing}, 
  title={Building Naturalistic Emotionally Balanced Speech Corpus by Retrieving Emotional Speech from Existing Podcast Recordings}, 
  year={2019},
  volume={10},
  number={4},
  pages={471-483},
  doi={10.1109/TAFFC.2017.2736999}}

@INPROCEEDINGS{yamauchi2023stylecapautomaticspeakingstylecaptioning,
  author={Yamauchi, Kazuki and Ijima, Yusuke and Saito, Yuki},
  booktitle={ICASSP 2024 - 2024 IEEE International Conference on Acoustics, Speech and Signal Processing (ICASSP)}, 
  title={STYLECAP: Automatic Speaking-Style Captioning from Speech Based on Speech and Language Self-Supervised Learning Models}, 
  year={2024},
  volume={},
  number={},
  pages={11261-11265},
  doi={10.1109/ICASSP48485.2024.10445977}}

@ARTICLE{chen2022wavlm,
  author={Chen, Sanyuan and Wang, Chengyi and Chen, Zhengyang and Wu, Yu and Liu, Shujie and Chen, Zhuo and Li, Jinyu and Kanda, Naoyuki and Yoshioka, Takuya and Xiao, Xiong and Wu, Jian and Zhou, Long and Ren, Shuo and Qian, Yanmin and Qian, Yao and Wu, Jian and Zeng, Michael and Yu, Xiangzhan and Wei, Furu},
  journal={IEEE Journal of Selected Topics in Signal Processing}, 
  title={WavLM: Large-Scale Self-Supervised Pre-Training for Full Stack Speech Processing}, 
  year={2022},
  volume={16},
  number={6},
  pages={1505-1518},
  doi={10.1109/JSTSP.2022.3188113}
}

@inproceedings{ma2023emotion2vecselfsupervisedpretrainingspeech,
    title = "emotion2vec: Self-Supervised Pre-Training for Speech Emotion Representation",
    author = "Ma, Ziyang  and
      Zheng, Zhisheng  and
      Ye, Jiaxin  and
      Li, Jinchao  and
      Gao, Zhifu  and
      Zhang, ShiLiang  and
      Chen, Xie",
    editor = "Ku, Lun-Wei  and
      Martins, Andre  and
      Srikumar, Vivek",
    booktitle = "Findings of the Association for Computational Linguistics: ACL 2024",
    month = aug,
    year = "2024",
    address = "Bangkok, Thailand",
    publisher = "Association for Computational Linguistics",
    url = "https://aclanthology.org/2024.findings-acl.931/",
    doi = "10.18653/v1/2024.findings-acl.931",
    pages = "15747--15760"
}

@misc{lacombe-etal-2024-parler-tts,
  author = {Yoach Lacombe and Vaibhav Srivastav and Sanchit Gandhi},
  title = {Parler-TTS},
  year = {2024},
  publisher = {GitHub},
  journal = {GitHub repository},
 url={https://github.com/huggingface/parler-tts}
}

@INPROCEEDINGS{guo2022promptttscontrollabletexttospeechtext,
  author={Guo, Zhifang and Leng, Yichong and Wu, Yihan and Zhao, Sheng and Tan, Xu},
  booktitle={ICASSP 2023 - 2023 IEEE International Conference on Acoustics, Speech and Signal Processing (ICASSP)}, 
  title={Prompttts: Controllable Text-To-Speech With Text Descriptions}, 
  year={2023},
  volume={},
  number={},
  pages={1-5},
  doi={10.1109/ICASSP49357.2023.10096285}}

@inproceedings{radford2021learning,
  title={Learning transferable visual models from natural language supervision},
  author={Radford, Alec and Kim, Jong Wook and Hallacy, Chris and Ramesh, Aditya and Goh, Gabriel and Agarwal, Sandhini and Sastry, Girish and Askell, Amanda and Mishkin, Pamela and Clark, Jack and others},
  booktitle={International conference on machine learning},
  pages={8748--8763},
  year={2021},
  organization={PMLR}
}

\end{document}